\documentclass[showpacs,preprintnumbers,amsmath,amssymb,preprint]{revtex4}

\usepackage{amsmath,amssymb,amsfonts,latexsym}

\newcommand{\be}{\begin{equation}}
\newcommand{\ee}{\end{equation}}
\newcommand{\bea}{\begin{eqnarray}}
\newcommand{\eea}{\end{eqnarray}}

\newcommand{\bas}{\begin{eqnarray*}}
\newcommand{\eas}{\end{eqnarray*}}
\newcommand{\ba}{\begin{array}}
\newcommand{\ea}{\end{array}}

\begin{document}

\title{Twisted Covariant Noncommutative Self-dual Gravity\footnote{
Dedicated to the memory of Julius Wess.}}
\author{S. Estrada-Jim\'enez$^{a}$}
\thanks{E-mail address: {\tt sestrada@unach.mx}}

\affiliation{Centro de Estudios en F\'{\i}sica y Matem\'aticas B\'asicas y Aplicadas,\\
Universidad Aut\'onoma de Chiapas\\
Calle 4$^a$ Oriente Norte. 1428, Tuxtla Guti\'errez, Chiapas,
M\'exico}

\author{H. Garc\'{\i}a-Compe\'an$^{b}$}
\thanks{E-mail address: {\tt compean@fis.cinvestav.mx}}
\affiliation{Departamento de F\'{\i}sica,
Centro de Investigaci\'on y de Estudios Avanzados del IPN\\
P.O. Box 14-740, 07000 M\'exico D.F., M\'exico}
\affiliation{Centro de Investigaci\'on y de Estudios Avanzados del
IPN, Unidad Monterrey, PIIT, V\'{\i}a del Conocimiento 201,
Autopista nueva al Aeropuerto km 9.5,  66600, Apodaca Nuevo
Le\'on, M\'exico}

\author{ O. Obreg\'on$^{c}$}
\thanks{E-mail address: {\tt octavio@fisica.ugto.mx}}

\affiliation{Instituto de F\'{\i}sica de la Universidad de Guanajuato\\
P.O. Box E-143, 37150, Le\'on Gto., M\'exico}

\author{C. Ram\'{\i}rez$^d$}

\thanks{E-mail address: {\tt cramirez@fcfm.buap.mx}}
\affiliation{Facultad de Ciencias F\'{\i}sico Matem\'aticas,
Universidad
Aut\'onoma de Puebla\\
P.O. Box 1364, 72000, Puebla, M\'exico}

\vskip -1truecm
\date{\today}

\begin{abstract}
A twisted covariant formulation of noncommutative self-dual
gravity is presented. The formulation for constructing twisted
noncommutative Yang-Mills theories is used. It is shown that the
noncommutative torsion is solved at any order of the
$\theta$-expansion in terms of the tetrad and some extra fields of
the theory. In the process the first order expansion in $\theta$
for the Pleba\'nski action is explicitly obtained.
\end{abstract}

\vskip -1truecm 

\maketitle

\vskip -1.3truecm
\newpage

\setcounter{equation}{0}
\section{Introduction}

Noncommutative structures in field theory have been studied over
the years and the subject has been introduced in several forms.
The idea of noncommutative space-time seems to be firstly proposed
by Heisenberg \cite{heisenberg}, as a possible cut-off to cure UV
divergences in quantum field theory (for some historical remarks,
see \cite{wesshisto}). This idea was further developed by H.S.
Snyder and applied it to find an implementation of the Lorentz
symmetry on a space-time with non-commuting coordinates
\cite{snyderone}. Snyder's construction was realized in
(4+1)-dimensional space-time, however it allows coordinate
transformations which break down the Lorentz symmetry in a
(3+1)-spacetime subspace. The extension to include the gauge
symmetry of the electromagnetic field was pursued in a second
paper \cite{snydertwo} (for a recent review, see
\cite{bietenholz}).

Deformation quantization and Connes approach are two of the most
used formulations of noncommutative spaces. On one hand,
deformation quantization was first introduced in the context of
phase-space quantization \cite{dq}, some few years before Snyder's
paper and finally formulated as an alternative quantization method
in Ref. \cite{BFFLS} (for a review, see \cite{zachosrev}). On the
other hand, Connes noncommutative geometry \cite{connes} is a
rigorous mathematical setting containing non-trivial structures
originated from von Neumann \cite{vN} and  Gelfand-Naimark results
\cite{GN}.

An important step done recently, was the discovery that noncommutative
gauge theory is obtained
naturally from non-perturbative string theory (D-brane physics)
via the Seiberg-Witten map \cite{Seiberg:1999vs}.
Furthermore, M-theory, in its M(atrix) theory approach,
was also shown to be compatible with these noncommutative
structures \cite{CDS}. This relation to string
theory has been one of the main motivations to further explore the
physics of noncommutative theories.

Non-commutative field theory can incorporate nonlocal effects in
field theory at the classical and quantum levels in an interesting
and subtle way. For instance it gives rise to surprising effects
like the IR/UV mixing \cite{minwalla} (for some reviews, see
\cite{ncreview,GraciaBondia:2001tr}). Recently non-perturbative
studies (via Monte Carlo simulations) seem to support the
existence of this mixing \cite{panero}.

Noncommutative field theories can be carried over to SU($N$) gauge
theories through the implementation of the Universal Enveloping
Algebra associated to the Lie algebra of the gauge group (of the
limiting commutative field theory). Consequently, the Standard
Model or GUT's models \cite{wesssm} can be constructed in this way
by using the Seiberg-Witten map. Similarly, noncommutative
versions of topological and self-dual gravity \cite{topo} can be
constructed by using the same methods.

In fact, there are numerous proposals of noncommutative gravity
theories in four dimensions \cite{ncgrav}. However, they do not
have a clear relation to string theory as the gauge theory
counterpart do \cite{Alvarez-Gaume:2006bn}. Moreover, the
diffeomorphism invariance turns out to be broken even at the
classical level. This is the problem of covariance and there is
evidence that noncommutative field theories also could be
non-unitary and violate causality (see for instance,
\cite{Seiberg:2000gc}). This of course has consequences for the
consistence of the theory.

More recently, proposals of a formulation of a covariant
noncommutative field theory have been made \cite{wess0,chaichian}
(see also, \cite{koch,oeckl}). In these proposals the Lorentz
symmetry transformations are deformed by a twist in order that the
noncommutative theory be invariant. By this twist the Leibniz rule
of the transformations on the Moyal product of two fields is
consistently deformed as
\begin{equation}
\delta^\star_\omega(\psi\star\phi)=\delta^\star_\omega\psi\star\phi+
\psi\star\delta^\star_\omega\phi,
\end{equation}
where $\delta^\star_\omega$ is a noncommutative variation operator
to be defined below. The twist has been formulated for
diffeomorphisms in \cite{wm}, in such a way that the algebraic
structure of the Lie algebra of vector fields on the manifold is
deformed into a noncommutative algebra of diffeomorphisms, keeping
the noncommutative parameter $\theta$ constant. This allows to
construct geometric composite covariant objects in the deformed
algebra, in particular metrics, covariant derivatives, curvature
and torsion. In this way in \cite{wessone} a twisted covariant
noncommutative Einstein-Hilbert action was given, which has been
further explored in \cite{wesstwo} (for some reviews on the
subject, see \cite{reviewwess}). In a similar spirit, gauge
symmetries can be twisted giving rise to covariant noncommutative
gauge theories described in
\cite{oeckl,Vassilevich:2006tc,Aschieri:2006ye} and further
developed in \cite{workonwess}. An interesting point of this
formulation is that the language of differential forms can be used
to obtain covariant results.

In \cite{Wess:2006zb}, J. Wess has given an explicit realization
of the twisted co-product (1) for gauge symmetry. This formulation
makes use of the functional calculus language from field theory,
which allows to explicitly restrict the transformations to the
fields, and to avoid the problem that the derivatives of the Moyal
product are not covariant. In the present paper we follow these
results, and generalize them to diffeomorphisms, in order to
construct a twisted covariant noncommutative formulation of
Plebanski's self-dual gravity. This involves simultaneously local
Lorentz and diffeomorphism transformations. As is well known,
Pleba\'nski's \cite{pleban} self-dual gravity is a topological
constrained $SL(2,\mathbb{C})$ $BF$ theory, and self-dual
variables have been the starting point to find loop variables to
quantize the gravitational field \cite{ashtekar}. In \cite{topo}
we have described $SL(2,\mathbb{C})$ noncommutative topological
and self-dual gravities, respectively. Though the theories are
manifestly Lorentz invariant, the diffeomorphism invariance
remains broken. In this respect they are not fully symmetric with
respect to the whole noncommutative symmetries. In this paper we
use the twisted formalism to give a noncommutative
$SL(2,\mathbb{C})$ $BF$ theory, {\it invariant} under twisted
local Lorentz and diffeomorphism transformations. In order to do
that, we exhibit a simple noncommutative version of the volume
form, given by the product of the one-form tetrads. Then we will
implement the noncommutative constraints in such a way that we get
a noncommutative version of Pleba\'nski action which is not only
invariant under twisted Lorentz transformations but also under
twisted diffeomorphism ones. We will show that the torsion can be
solved at any order of an expansion on $\theta$. The Lagrangian,
the torsion and other relevant expressions are explicitly
calculated in the present paper.

This paper is organized as follows, in Section II we give an
overview of twisted covariant non-commutative gauge and gravity
theories. In the process we prove that the functional derivative
methods introduced in Ref. \cite{Wess:2006zb} for gauge fields,
can be carried over to the construction of noncommutative
gravitational fields with twisted symmetries (a basic detailed
calculation is summarized in the appendix). In Section III we give
an overview of self-dual gravity. In section IV we construct the
twisted covariant self-dual gravity. Section V is devoted to the
final remarks.

\bigskip
\section{Noncommutative Gauge and Gravity Theories Constructed Via Twisting}

In the present section we describe some important features of the
twisted gauge and diffeomorphism transformations, that will be
necessary in Sec. IV.

Noncommutativity is introduced ordinarily through a Moyal-Weyl
space-time, with commutation relations given by (for recent
reviews see, \cite{ncreview})
\begin{equation} [\widehat x^\mu,\widehat x^\nu] =
i \theta^{\mu\nu},
\end{equation}
where the antisymmetric matrix $\theta^{\mu \nu}$ has constant
entries. This noncommutativity can be realized through the
Wigner-Moyal correspondence, by means of the Moyal product
\begin{equation}
f\star g(x) =\mu_\star(f\otimes g)(x)=  \exp\bigg\{ {i \over 2}
\theta^{\mu \nu} \frac{\partial}{\partial
y^\mu}\frac{\partial}{\partial z^\nu} \bigg\}
f(y)g(z)\bigg|_{x=y=z},\label{moyal}
\end{equation}
where $\mu_\star = \mu \circ {\cal F}^{-1}$, $\mu$ is the product
map $\mu(f \otimes g)= f g$, with $ {\cal F}=
e^{-\frac{i}{2}\theta^{\mu\nu}\partial_\mu \otimes \partial_\nu}$.
When necessary, we will write $\mu(f \otimes g) =f\cdot g$, in
order to stress that we are speaking of the commutative product.
The presence of $\theta^{\mu \nu}$ as a constant matrix and of
ordinary derivatives in this definition leads to the loss of
covariance. In particular it has been shown that Lorentz
invariance is violated (see for instance, \cite{Seiberg:2000gc})
and there are problems to include in a theory representations with
different charges \cite{wesssm}.

If we consider a gauge group of transformations, one can construct
a covariant deformed theory by introducing an appropriate {\it
twisting} of the transformation law of the Moyal product of fields
in some specific representation ${\cal R}$. The ingredients are:
$(i)$ a Lie algebra ${\cal G}$, $(ii)$ an action of the Lie
algebra ${\cal G}$ on the space of functions ${\cal A}=Fun(M)$ of
the space-time manifold $M$ that one wants to deform into a
noncommutative algebra ${\cal A}_\theta$, and $(iii)$ a twist
element ${\cal F}$, constructed with the generators of the Lie
algebra ${\cal G}$. Then by twisting the gauge and diffeomorphism
Lie algebras we can obtain a covariant noncommutative theory of
gauge fields or gravitational fields, respectively.

We start by considering a gauge group $G$, with a Lie algebra
${\cal G}$ in some irreducible representation ${\cal R}$. The
symmetry properties of the field theory on spacetime $M$ can be
lifted in a natural way to the universal enveloping algebra

$U({\cal G},{\cal R})$ of ${\cal G}$ in the irrep ${\cal R}$. This
has a structure of {\it Hopf algebra} $(U({\cal G},{\cal
R}),m,e;\Delta,\varepsilon;S)$ where $\Delta: U({\cal G},{\cal R})
\to U({\cal G},{\cal R}) \otimes U({\cal G},{\cal R})$ is the
co-product (for a more detailed description of the Hopf algebra
structure the reader can consult for instance, \cite{pressley}).
In the commutative theory the transformation law of the product of
two fields, i.e. the Leibniz rule, is given by
\begin{equation}
\delta_\alpha(\phi \cdot \psi)= \mu
\big[\Delta(\delta_\alpha)(\phi\otimes\psi)\big]=
(\delta_\alpha\phi) \cdot \psi + \phi \cdot (\delta_\alpha \psi),
\label{trece}
\end{equation}
where $\Delta(\delta_\alpha)= \delta_\alpha\otimes
1+1\otimes\delta_\alpha$ is the co-product.  In the noncommutative
theory, the definition of the co-product is more elaborated.
Indeed, the Moyal product of fields involves non-covariant
derivatives, hence we would expect that it will have a complicated
nonlinear transformation. However the coproduct can be generalized
by considering a twisting of Eq. (\ref{trece}).

\vskip 1.5truecm
\noindent
{\it Twisted Gauge Transformations}

Let us write the transformation of a matter field as
$\delta_\alpha\phi_j(x)=i\alpha^l(x)[T_l\phi(x)]_j=S_\alpha\phi_j(x)$,
where $S_\alpha$ is given by \cite{Wess:2006zb}
\begin{equation}
S^\phi_\alpha=i\int{dz
\alpha^l(z)[T_l\phi(z)]_j\frac{\delta}{\delta\phi_j}}.\label{g1}
\end{equation}
Then we define the noncommutative infinitesimal transformations by
\begin{equation}
\delta_\alpha^\star\phi=S^\star_\alpha\phi=
\delta_\alpha\phi=S_\alpha\phi, \label{ese}
\end{equation}
which, according to \cite{wessone} it can be written also as
$\delta_\alpha^\star\phi= - X_\alpha^\star\star\phi$. Here we will
follow the formulation (\ref{ese}), which makes explicit that the
transformations act only on the fields of the theory. Thus we can
define a restricted Moyal product which operates only on the
fields
\begin{equation}
\mu_\star(\phi\otimes\psi)= \mu \circ {\cal
F}^{-1}(\phi\otimes\psi)=\phi\star\psi,
\end{equation}
where ${\cal F}$ is a bilinear functional operator which acts on
all the fields of the theory, here denoted as $\{\phi_k\}$
\begin{equation}
{\cal
F}=e^{-\frac{i}{2}\theta^{\mu\nu}\int{dz\partial_\mu\phi_k(z)\frac{\delta}{\delta
\phi_k(z)}}
\otimes\int{dy\partial_\nu\phi_l(y)\frac{\delta}{\delta
\phi_l(y)}}}
\label{twist}
\end{equation}
and $\phi$, $\psi \in {\cal A}_\theta$.

This restricted Moyal product does not act on functions like the
parameters of the symmetry transformations, i.e., if $f$ is a
function not related to the fields of the theory, then
$\mu_\star(f\otimes \Psi)=\mu(f\otimes \Psi)=f\Psi$.

According to the twisted noncommutative theories, the Leibniz rule
is written in terms of a twisted co-product of (\ref{ese})
$$
\delta_\alpha^\star(\phi\star\psi) =\mu_\star[\Delta_{\cal
F}(S_\alpha)(\phi\otimes\psi)]
$$
\begin{equation}
= (\delta_\alpha^\star \phi)\star\psi + \phi \star
(\delta_\alpha^\star\psi), \label{coproduct1}
\end{equation}
where
\begin{equation}
\Delta_{\theta}(S_\alpha) \equiv \Delta_{\cal F}(S_\alpha)={\cal
F}^{-1}\Delta(S_\alpha){\cal F}
\end{equation}
is the Drinfeld's twisted co-product arising in the definition of
quasi-triangular Hopf algebras \cite{drinfeld,pressley} and
$\Delta(S_\alpha)= S_\alpha \otimes 1+1 \otimes S_\alpha$ is the
commutative co-product of (\ref{g1}). It can be shown in a
straightforward, although somewhat cumbersome way, that the
co-product (\ref{coproduct1}) gives the same result as in
\cite{Wess:2006zb} (an explicit derivation is worked out in the
appendix)
\begin{equation}
\delta^\star_\alpha(\phi_r\star\phi_s)=i\alpha^l\cdot\left[(T_l\phi)_r
\star\phi_s+\phi_r\star(T_l\phi)_s\right].
\end{equation}
The right hand side looks quite similar to the usual Leibniz rule,
but is radically different because the Moyal product does not act
on the transformation parameters, as they multiply the rest of the
expression to their right with the ordinary commutative
multiplication. The reason of why to use such a complicated
expression like (\ref{coproduct1}), is that it is part of a Hopf
algebra \cite{wesstwo} which ensures its consistency, e.g. the
associativity.

\vskip 1.5truecm
\noindent
{\it Twisted Gauge Fields}

Things work quite similar for gauge fields $A^l_\mu$. In this case
the action of the transformations is given by \cite{Wess:2006zb}
\begin{equation}
S^A_\alpha=i\int{dz\left[\partial_\mu\alpha^l(z)-
\alpha^r(z){f_{rs}}^l A_\mu^s(z)\right]\frac{\delta}{\delta
A_\mu^l(z)}}.
\end{equation}
For example, the transformation rule of the product
$A_\mu\star\phi$ is given by the expression (\ref{coproduct1})
\begin{equation}
\delta_\alpha^\star(A_\mu\star\phi)=\mu_\star[\Delta_{\cal
F}(S_\alpha)(A_\mu \otimes\phi)]= \mu_\star[{\cal
F}^{-1}\Delta(S_\alpha){\cal
F}(A_\mu\otimes\phi)]=\partial_\mu\alpha\cdot\phi+i\alpha\cdot(A_\mu\star\phi),
\end{equation}
where it has been taken into account that the fields are in
different representations of the gauge group,
$\Delta(S_\alpha)=S^A_\alpha\otimes 1+1\otimes S^\phi_\alpha$.
Such expressions containing different fields can be handled
considering, as in the case of the functional operator $\cal F$,
that the transformation $S_\alpha$ must contain a sum over all the
fields of the theory. Thus, the covariant derivative
\begin{equation}
D^\star_\mu\phi=\partial_\mu\phi-iA_\mu\star\phi,
\end{equation}
fulfils $\delta^\star_\alpha D^\star_\mu\phi=i\alpha\cdot
D^\star_\mu\phi$. The field strength is obtained as usual
\begin{equation}
\left(D^\star_\mu\star D^\star_\nu-D^\star_\nu\star
D^\star_\mu\right)\phi =-i\left(\partial_\mu A_\nu-\partial_\nu
A_\mu-i[A_\mu\stackrel{\star}{,} A_\nu]\right)\star\phi =-i
F_{\mu\nu}^\star\star\phi.
\end{equation}
In order to get its transformation rule,
we need to compute the transformation of $A_\mu\star A_\nu$, which turns out to be
\begin{equation}
\delta_\alpha^\star(A_\mu\star A_\nu)=
\mu_\star[{\cal F}^{-1}\Delta(S_\alpha){\cal F}(A_\mu\otimes A_\nu)]
=\partial_\mu\alpha\cdot A_\nu+\partial_\nu\alpha\cdot A_\mu+i\alpha^l\cdot
[T_l,A_\mu\star A_\nu].
\end{equation}
Thus the field strength transforms as usual, $\delta_\alpha^\star
F_{\mu\nu}^\star =i\alpha^l\cdot [T_l,F_{\mu\nu}^\star].$ In order
to ensure that the covariant derivative has the correct
properties, it must have a co-product. Let us write the covariant
derivative as
\begin{equation}
D^\star_\mu=\int{dz\left[\partial_\mu\phi_l(z)- i{A_{\mu
l}}^k(z)\star\phi_k(z)\right]\frac{\delta}{\delta \phi_l(z)}}.
\end{equation}
Then we have, after straightforward but laborious computations
\begin{equation}
D^\star_\alpha (\phi_r \star \phi_s) = \mu_\star[{\cal
F}^{-1}\Delta(D_\mu^\star){\cal F}(\phi_r\otimes\phi_s)]=
(D_\mu^\star\phi)_r\star\phi_s+\phi_r\star(D_\mu^\star\phi)_s,
\label{coproductD}
\end{equation}
where $\Delta(D_\mu^\star)=D_\mu^\star\otimes 1+1\otimes
D_\mu^\star$. For the adjoint representation we have similar
rules.

\vskip 1.5truecm
\noindent {\it Twisted Diffeomorphisms}

For diffeomorphisms it is not obvious that if we use Drinfeld's
twisted coproduct, we will get similar results. Let us consider
for instance a covariant vector field $U_\mu$
\begin{equation}
\delta_\xi U_\mu=-\xi^\nu\partial_\nu U_\mu-\partial_\mu\xi^\nu
U_\nu.\label{vector}
\end{equation}
As far as the second term in the r.h.s. is a matrix
transformation, the considerations of the preceding section can be
applied to it. Thus, in order to see if the previous formulation
can be applied here, it is enough to consider scalar fields
$\phi$, transforming as: $\delta_\xi
\phi=-\xi^\nu\partial_\nu\phi= S^\phi_\xi\phi$. Here we define
\begin{equation}
S^\phi_\xi=-\int{dz
\xi^\mu(z)\partial_\mu\phi(z)\frac{\delta}{\delta\phi(z)}}.\label{tr1}
\end{equation}
Thus, if we apply (\ref{coproduct1}), after computations (the
detailed calculation in the appendix A) we get
\begin{equation}
\delta_\xi^\star(\phi\star\psi)=\mu_\star[\Delta_{\cal
F}(S_\xi)(\phi\otimes\psi)]=-\xi^\mu\cdot\left(\partial_\mu\phi\star\psi+
\phi\star\partial_\mu\psi\right).
\end{equation}
Hence, for a covariant vector field we have
\begin{equation}
S^{U_\mu}_\xi=-\int{dz \left[\xi^\nu(z)\partial_\nu U_\rho(z)+
\partial_\rho\xi^\nu(z) U_\nu(z)\right]\frac{\delta}{\delta
U_\rho(z)}}.
\label{cvf}
\end{equation}
Similarly, the procedure can be also carried over for a
contravariant vector field $V^\mu$. Therefore, we can compute the
transformations of mixed products like
\begin{eqnarray}
\delta_\xi^\star(\phi\star U_\mu)&=&\mu_\star[\Delta_{\cal
F}(S_\xi)(\phi\otimes U_\mu)]=
\mu_\star[{\cal F}^{-1}(S^\phi_\xi\otimes 1
+1\otimes S^{U_\mu}_\xi){\cal F}(\phi\otimes U_\mu)]\nonumber\\
&=&-\xi^\nu\cdot\left(\partial_\nu\phi\star U_\mu+ \phi\star
\partial_\nu U_\mu\right)-\partial_\mu\xi^\nu\cdot (\phi\star U_\nu)
\end{eqnarray}
or
\begin{eqnarray}
\delta_\xi^\star(V^\mu\star U_\nu)&=&\mu_\star[\Delta_{\cal
F}(S_\xi)(V^\mu\otimes U_\nu)]=
\mu_\star[{\cal F}^{-1}(S^{V^\mu}_\xi\otimes 1+1\otimes S^{U_\nu}_\xi){\cal F}(V^\mu\otimes U_\nu)]
\nonumber\\
&=&-\xi^\rho\cdot\left(\partial_\rho V^\mu\star U_\nu +V^\mu\star
\partial_\rho U_\nu\right) + \partial_\rho\xi^\mu\cdot
(V^\rho\star U_\nu) -\partial_\nu\xi^\rho\cdot (V^\mu\star
U_\rho).
\end{eqnarray}
Thus the Moyal product of tensor quantities lead to higher order
tensors as usual, and the contraction of indices of tensor
quantities lead to lower order tensors, for instance
\begin{equation}
\delta_\xi^\star(V^\mu\star U_\mu)=-\xi^\nu\cdot\partial_\nu
\left(V^\mu\star U_\mu\right).
\end{equation}

\vskip 1.5truecm
\noindent
{\it Twisted Differential Forms}

The above properties allow us to define in the usual way
differential 1-forms $U=U_\mu dx^\mu$, which can be extended to
higher order differential forms if the differentials $dx$ behave
as constants under the Moyal product, i.e. for the product of two
1-differential forms $U$ and $V$ we have
\begin{equation}
U\stackrel{\star}{\wedge} V=U_\mu dx^\mu\stackrel{\star}{\wedge}
V_\nu dx^\nu= (U_\mu\star V_\nu) dx^\mu\wedge dx^\nu.
\end{equation}
The essential point is the noncommutative exterior derivative
which is defined as commutation of the following diagram:
\begin{eqnarray}
{f\in {\cal A}}&\stackrel{\cal W}{\longmapsto}&{f\in
{\cal{A}}_\theta} \nonumber\\
d \downarrow &&\downarrow d^\star \nonumber\\
{(df) \in \Lambda^1({\cal A})}&\stackrel{\cal W}{\longmapsto}&
{(d^\star \stackrel{\star}{\triangleright} f) \in
\Lambda^1({\cal{A}}_\theta)}\nonumber
\end{eqnarray}
\begin{equation}
(d^\star \stackrel{\star}{\triangleright} f) = (\partial^\star_\mu
\stackrel{\star}{\triangleright} f) dx^\mu = (\partial_\mu f)
dx^\mu =(df).
\end{equation}
Here $\Lambda^1({\cal{A}}_\theta)$ is a left (or right) module
over ${\cal{A}}_\theta$, i.e. an ${\cal{A}}_\theta$-module. In the
diagram ${\cal W}$ is the map given by the Weyl-Wigner-Moyal
correspondence \cite{dq}, which is an isomorphism. With this
definition it easy to show that
\begin{equation}
d^\star \stackrel{\star}{\triangleright} d^\star
\stackrel{\star}{\triangleright} f= 0,
\end{equation}
for any $f$. Thus $d^\star \stackrel{\star}{\triangleright}
d^\star=0.$ In this way, the differential of a differential form
gives, as usual, higher order differential forms
\begin{equation}
dU=\partial_\mu U_\nu dx^\nu\wedge dx^\mu.
\end{equation}
For $p$-forms
\begin{eqnarray}
{U_p\in \Lambda^p({\cal A})}&\stackrel{\cal W}{\longmapsto}&{U_p
\in
\Lambda^p({\cal{A}}_\theta}) \nonumber\\
d_p \downarrow &&\downarrow d^\star_p \nonumber\\
({d_p U_p) \in \Lambda^{p+1}({\cal A})}&\stackrel{\cal
W}{\longmapsto}& {(d^\star_p \stackrel{\star}{\triangleright} U_p)
\in \Lambda^{p+1}({\cal{A}}_\theta)}, \nonumber
\end{eqnarray}
such that the diagram commutes, i.e.,
\begin{equation}
(d^\star_p \stackrel{\star}{\triangleright} U_p)= (d_p U_p).
\end{equation}
Here $\Lambda^p({\cal A}_\theta)$ is also a
${\cal{A}}_\theta$-module.

In the general case for the wedge product of a $p$-form $U_p$ by a
$q$-form $V_q$, the usual graded Leibniz rule is satisfied
\begin{equation}
d^\star \stackrel{\star}{\triangleright}
(U_p\stackrel{\star}{\wedge} V_q)=(d^\star
\stackrel{\star}{\triangleright} U_p) \stackrel{\star}{\wedge}
V_q+(-1)^{p}U_p\stackrel{\star}{\wedge} (d^\star
\stackrel{\star}{\triangleright}V_q).
\end{equation}

Under this scheme, in which transformations $S_\xi$ act only on
the fields, differential forms will not transform under
diffeomorphisms as scalar fields. Usually, the transformation of
the field is compensated by the coordinate transformation. In the
present case, considering for instance a one-form $U$, we have
$\delta^\star_\xi U=dx^\mu S_\xi U_\mu$, where $S_\xi U_\mu$ is
given by (\ref{cvf}). However the interesting feature is that,
despite of this undesirable property, four-forms continue to
transform (in four dimensions) as invariant densities. Indeed, let
us consider $U=dx^\mu\wedge dx^\nu\wedge dx^\rho\wedge dx^\sigma
U_{\mu\nu\rho\sigma}=
dV\varepsilon^{\mu\nu\rho\sigma}U_{\mu\nu\rho\sigma}$, where
$\varepsilon^{\mu\nu\rho\sigma}$ is the Levi-Civita symbol in four
dimensions. Then we have $\delta^\star_\xi
U=dV\varepsilon^{\mu\nu\rho\sigma}S_\xi U_{\mu\nu\rho\sigma}$,
that is
\begin{eqnarray}
\delta^\star_\xi U &=&dV\varepsilon^{\mu\nu\rho\sigma}\left(
-\xi^\lambda\partial_\lambda U_{\mu\nu\rho\sigma}
-\partial_\mu\xi^\lambda U_{\lambda\nu\rho\sigma}
-\partial_\nu\xi^\lambda U_{\mu\lambda\rho\sigma}
-\partial_\rho\xi^\lambda U_{\mu\nu\lambda\sigma}
-\partial_\sigma\xi^\lambda U_{\mu\nu\rho\lambda}\right)\nonumber\\
&=&dV\varepsilon^{\mu\nu\rho\sigma}\left(
-\xi^\lambda\partial_\lambda U_{\mu\nu\rho\sigma}
-\partial_\lambda\xi^\lambda U_{\mu\nu\rho\sigma}\right)
=-\partial_\lambda\left(\xi^\lambda U\right),
\end{eqnarray}
where we used the identity:
$\varepsilon^{\mu\nu\rho\sigma}\xi^\lambda+\ {\rm cyclic\
permutations \ of} \ \{\mu\nu\rho\sigma\lambda\}\equiv0$. This
result can be understood from the invariance of the action under
the transformations of the fields, as can be seen from
\begin{eqnarray}
\delta A&=&\int d^4 x' {\cal L}[\phi'(x'),\partial_\mu'\phi'(x')]-
\int d^4 x {\cal L}[\phi(x),\partial_\mu\phi(x)]\nonumber\\
&=&\int d^4 x {\cal L}[\phi'(x),\partial_\mu\phi'(x)]- \int d^4 x
{\cal L}[\phi(x),\partial_\mu\phi(x)]=0.
\end{eqnarray}
In this way we can construct noncommutative invariant actions by
means of four forms. For instance, if we consider the product of
four one-form tetrad, we get
\begin{equation}
e^a\stackrel{\star}{\wedge}e^b\stackrel{\star}{\wedge}e^c
\stackrel{\star}{\wedge}e^d=
dV\varepsilon^{\mu\nu\rho\sigma}e_\mu^{\ a}\star e_\nu^{\ b} \star
e_\rho^{\ c}\star e_\sigma^{\ d}, \label{volumeone}
\end{equation}
which is not antisymmetric in the indices $a,b,c,d$ and
consequently does not give the determinant of the tetrad. However
it is still an invariant density, and we must take care only about
Lorentz invariance. For example, by contracting this quantity with
a suitable four tensor, we get an expression invariant under
twisted Lorentz plus diffeomorphisms transformations, as follows
\begin{equation}
(\delta_\Lambda +
\delta_\xi)(e^a\stackrel{\star}{\wedge}e^b\stackrel{\star}{\wedge}e^c
\stackrel{\star}{\wedge}e^d\star
V_{abcd})=-\partial_\lambda[\xi^\lambda\cdot(e^a\stackrel{\star}
{\wedge}e^b\stackrel{\star}{\wedge}e^c
\stackrel{\star}{\wedge}e^d\star V_{abcd})]. \label{volumetwo}
\end{equation}
In this paper we consider the Plebanski's action of gravity
\cite{pleban}, which is a $BF$ theory (with constraints) with the
Lagrangian given by a four form. It will be twisted in the same
sense as a gauge theory and will be covariant under twisted
Lorentz and diffeomorphism transformations.

\section{Brief Overview of the Self-dual Formulation of Gravity}

In this section we overview the self-dual formulation of gravity
in four dimensions. We will follow Pleba\'nski paper
\cite{pleban}. Let start by considering a $SO(3)$ complex
connection one-form $\Omega=\Omega_iT^i$ $(i=1,2,3)$, with its
corresponding field strength $F=d\Omega+\Omega\wedge\Omega$ and
the $BF$-action
\begin{equation}
I=-4i\int {\rm Tr} \big(B\wedge F \big),\label{pleba}
\end{equation}
where $B$ is a Lie algebra valued two-form. Let us now write the
fields into their real and imaginary parts,
$\Omega=\frac{1}{2}(\omega+i\widetilde\omega)$,
$B=\frac{1}{2}(\Sigma+i\widetilde\Sigma)$ and
$F=\frac{1}{2}(R+i\widetilde{R})$. Now let us define
${\omega}^{i0}=-{\omega}^{0i}=-{\omega}^i$, ${\omega}^{00}=0$ and
$\omega^{ij}={\varepsilon^{ij}}_k\widetilde\omega^k$. Similarly
${R}^{0i}=-{R}^{i0}={R}^i$, ${R}^{00}=0$ and ${R}^{ij}=
{\varepsilon^{ij}}_k{\widetilde R}^k$. In this case, putting
things toghether, we get the $SO(3,1)$ field strength,
${R}^{ab}=d{\omega}^{ab}+{\omega}^{ac}\wedge{\omega_{c}}^b$ (with
$a,b=0,1,2,3$).

In general if $v^i$ is a complex $SO(3)$ field, its decomposition
into real and imaginary parts can be rewritten as a $SO(3,1)$

algebra valued self-dual field,
$v^i=\frac{1}{2}(v^{0i}-\frac{i}{2}{\varepsilon^{0i}}_{jk}v^{jk})=v^{(+)0i}$,
i.e. it fulfils ${\varepsilon^{ab}}_{cd}v^{(+)cd}=2iv^{(+)ab}$.
Moreover $u^iv_i=-\frac{1}{4}u^{(+)ab}v^{(+)}_{\ ab}$. The field
strength satisfies $R^{(+)ab}(\omega^{(+)})=R^{ab}(\omega^{(+)})$.

Following Pleba\'nski \cite{pleban}, the solution of the
constraints for the $B^i$ field is given by $\Sigma^{ab}=e^a\wedge
e^b$, where the tetrad $e^a$ are real one-forms, which are defined
up to a $SO(3,1)$ transformation. Thus, the action (\ref{pleba})
can be rewritten as
\begin{equation}\label{accionsd1}
I=i\int \Sigma^{(+)ab}\wedge R_{ab}^{(+)}=
\frac{1}{2}\int\left(\frac{1}{2}\varepsilon_{abcd}\, e^a\wedge
e^b\wedge R^{cd}+i\,e^a\wedge e^b\wedge R_{ab}\right).
\end{equation}
From the tetrad we obtain the torsion two-form:
$T^a=De^a=de^a+\omega^a_{\ b}\wedge e^b$, which satisfies the
Bianchi identity $DT^a=dT^a+\omega^a_{\ b}\wedge T^b\equiv R^a_{\
b}\wedge e^b$. Therefore, the second term in (\ref{accionsd1}) can
be written in terms of the torsion. Furthermore, if we write
$R^{ab}=R^{\mu\nu}e_\mu^{\ a}e_\nu^{\ b}$, the action
(\ref{accionsd1}) can be rewritten as
\begin{equation}
I=-\int {\rm det}e\,R_{\mu\nu}^{\ \ \mu\nu}(\omega)+
\frac{i}{2}\int e^a\wedge DT_a.
\label{acciondos}
\end{equation}
The first term is the Palatini action, with the tetrad $e$ and the
connection $\omega$ being independent fields. The variation of
$\omega$ on this term gives
\begin{equation}
\delta_\omega I=i\delta_\omega\int\Sigma^{ab} \wedge
R_{ab}\left(\omega^{(+)}\right)= 2i\int e^a\wedge  T^b\wedge
\delta\omega^{(+)}_{ab}=0.
\end{equation}
This is a complex equation, where the coefficient $e^a \wedge T^b$
is real. Therefore, if we set to zero the real and the imaginary
parts separately, we get the equation $e^a\wedge T^b-e^b\wedge
T^a=0$, from which turns out that the torsion vanishes,
$T_{\mu\nu}^{\ \ a}=0$, with the well known solution given by the
second Cartan structure equation
\begin{equation}
\omega_{\mu\nu\rho}=\frac{1}{2}\left[e_{\mu a}(\partial_\nu e_{ \
\rho}^a-\partial_\rho e_{ \ \nu}^a)- e_{\nu a}(\partial_\rho e_{\
\mu}^a-\partial_\mu e_{ \ \rho}^a)- e_{\rho a}(\partial_\mu e_{ \
\nu}^a-\partial_\nu e_{ \ \mu}^a)\right].\label{conexion}
\end{equation}
If we put it back into the action (\ref{acciondos}), we get the
Einstein-Hilbert action
\begin{equation}
I=\int\det e \: \cdot {R_{\mu\nu}}^{\mu\nu}d^4x.\label{accion1}
\end{equation}

\section{Twisted Covariant Noncommutative Self-dual Gravity}

Let us consider now the noncommutative theory of the action
(\ref{pleba}). In order to take into account the form of the
noncommutative field strength, we must extend the fields to the
universal enveloping algebra (UEA) of $su(2)$, given in this case
by $u(2)$.  The formulation of twisted gauge transformations
closes for arbitrary gauge groups \cite{Vassilevich:2006tc,
Aschieri:2006ye}. Thus, our proposal would be valid also for the
Lie algebra $su(2)$. However, as it was discussed in Ref.
\cite{Aschieri:2006ye}, the consistency of the classical equations
of motion of noncommutative Yang-Mills theory requires the use of
the associated UEA. In Appendix B we show that these
considerations apply also for BF actions. As we precisely need to
work out the equations of motion consistently (including the
torsion) from the noncommutative action corresponding to
(\ref{pleba}), we shall use the UEA of $su(2)$, namely $u(2)$.
Hence the action is now
\begin{equation}
A=-2i {\rm Tr}\int{\Sigma}\wedge\widehat{R} = -4i\int
\left(\Sigma^i\wedge\widehat{R}_i+\Sigma^4\wedge\widehat{R}_4\right),
\label{accion}
\end{equation}
where $\widehat{R}=d\omega+\omega\stackrel{\star}{\wedge}\omega$
is the noncommutative field strength and $T^A=
\{\sigma^i,\sigma^4={\bf 1}\}$, are the generators of the $u(2)$
algebra, which satisfy:
\begin{eqnarray}
&&[\sigma^i,\sigma^j]=2i\varepsilon^{ij}_{~~k}T^k, \quad
[\sigma^i,\sigma^4]=0, \quad \{\sigma^i,\sigma^j\}=2\delta^{ij}\sigma^4, \nonumber\\
&&\{\sigma^i,\sigma^4\}=2\sigma^i,\quad
\{\sigma^4,\sigma^4\}=2\sigma^4, \quad {\rm Tr}(\sigma^A\sigma^B)=
2\delta^{AB},
\end{eqnarray}
where $A,B=1,2,3,4.$ Following Sec. II, in particular formula
(\ref{volumetwo}), one can see that the action (\ref{accion}) is
invariant under twisted gauge and diffeomorphism transformations.

The field strength is given by $\widehat{R}=
\widehat{R}^iT_i+\widehat{R}^4T_4$, where
\begin{eqnarray}
\widehat{R}^i&=&d\omega^i + i \varepsilon_{~jk}^{i}\omega^j \stackrel{\star}{\wedge} \omega^k +
\omega^i \stackrel{\star}{\wedge} \omega^4 + \omega^4 \stackrel{\star}{\wedge} \omega^i, \\
\widehat{R}^4&=& d\omega^4 + \omega^i
\stackrel{\star}{\wedge}\omega_i+\omega^4
\stackrel{\star}{\wedge}\omega^4.
\end{eqnarray}
The first term (zero-th order) in the $\theta$-expansion of the
curvatures coincide with the commutative ones.

In terms of $SO(1,3)$ self-dual fields, by means of the relations
shown in the preceding section, the action (\ref{accion}) is given
by
\begin{eqnarray}
A&=& i \int \left[\Sigma^{(+)ab}\stackrel{\star}{\wedge} \left( d\omega_{ab}^{(+)}-
\omega_a^{(+)c} \stackrel{\star}{\wedge} \omega_{cb}^{(+)} +\omega_{ab}^{(+)} \stackrel{\star}{\wedge}
\omega^4+\omega^4 \stackrel{\star}{\wedge} \omega_{ab}^{(+)} \right)\right. \nonumber \\
&& \left.- 4\Sigma^4 \stackrel{\star}{\wedge}
\left(d\omega^4-\frac{1}{4}\omega^{(+)ab} \stackrel{\star}{\wedge}
\omega_{ab}^{(+)}+\omega^4 \stackrel{\star}{\wedge}
\omega^4\right) \label{ncaccion} \right],
\end{eqnarray}
where $\Sigma^{ab}$ and $\omega^{ab}$, which arise from $\Sigma^{i}$ and $\omega^{i}$, are real
and antisymmetric by construction, as in the commutative case.

This action must be written explicitly in terms of the tetrad in
order to be compared with the Einstein-Hilbert action. In the
commutative case, Pleba\'nski has formulated constraints on
$\Sigma^i$, whose solution is given by $\Sigma^{ab}=e^a\wedge
e^b$. After substitution of this solution into the commutative
action, the Palatini action turns out. Here we have a
noncommutative action, with an explicit dependence on the
noncommutativity parameter $\theta$. Thus the solution of the
equations of motion will be given by generic fields $\phi$'s
depending on $\theta$. This dependence can be made explicit by a
series expansion
\begin{equation}
\phi(\theta)=\sum_{n=0}^\infty\theta^{a_1b_1}\cdots\theta^{a_nb_n}
\phi^{(n)}_{a_1b_1\cdots a_nb_n}. \label{serie}
\end{equation}
Here we will make an ansatz on the form of the
dependence of $\Sigma^{ab}$ on $\theta$, given in terms of the
tetrad by
\begin{equation}
\Sigma^{ab}=\frac{1}{2}(e^a\stackrel{\star}{\wedge} e^b-
e^b\stackrel{\star}{\wedge} e^a).\label{sigma}
\end{equation}
It is easy to see that the power series dependence on $\theta$ of
this expression has only even powers. The next step is the
variation of the action (\ref{ncaccion}) with respect to the
connection $\omega^{(+)}_{ab}$, which gives us
\begin{equation}
\delta_{\omega^{(+)}}A=i \int \bigg\{ -d\Sigma^{ab}+
[\omega^{}\stackrel{\star}{,}\Sigma^{}]^{ab}-
[\omega^{4}\stackrel{\star}{,}\Sigma^{ab}] -
[\omega^{ab}\stackrel{\star}{,}\Sigma^{4}] \bigg\}\wedge
\delta\omega_{ab}^{(+)}=0,\label{ecs}
\end{equation}
where $[\omega^{}\stackrel{\star}{,}\Sigma^{}]^{ab}=
[\omega^{ac}\stackrel{\star}{,}\Sigma_c^{ \ b}] = (\omega
\stackrel{\star}{\wedge} \Sigma)^{ab} -  (\Sigma
\stackrel{\star}{\wedge} \omega)^{ab}.$ In order to see which are
the equations of motion arising from this variation, we take into
account that we are dealing with complex quantities. Let us
consider generic equations of the form
$E^{ab}\wedge\delta\omega^{(+)}_{ab}=E^{ab(+)}\wedge\delta\omega^{(+)}_{ab}=0$.
If $E^{ab}$ are real, the real and imaginary parts of the
equations give $E^{ab}=0$. If $E^{ab}$ are complex, by the
properties of the self-dual projector we can see that it is enough
to set to zero their real, or their imaginary parts. Indeed,
taking into account that
$\overline{E^{ab(+)}}=\overline{E^{ab}}^{(-)}$, where
$\overline{E}$ stands for the complex conjugated of $E$, we get
for the real or the imaginary parts the same result
\begin{equation}
\left(E^{ab(+)}\pm\overline{E^{ab}}^{(-)}\right)\wedge\delta\omega^{(+)}_{ab}=
E^{ab(+)}\wedge\delta\omega^{(+)}_{ab}=E^{ab}\wedge\delta\omega^{(+)}_{ab}=0.
\end{equation}
Hence if the real part of $E$ vanishes, then the imaginary one
vanishes as well. Further, if $f$ and $g$ are real functions, then
$\overline{f\star g}=g\star f$. Thus, from the real part of the
coefficient of $\delta \omega^{(+)}_{ab}$ in (\ref{ecs}), we
obtain the equations of motion
\begin{equation}
2d\Sigma^{ab}-
[\omega\stackrel{\star}{,}\Sigma]^{ab}+[\omega\stackrel{\star}{,}\Sigma]^{ba}-2i\left\{
[\eta_{2}\stackrel{\star}{,}\Sigma^{ab}]
-[\omega^{ab}\stackrel{\star}{,}\lambda_{2}]
-\frac{1}{2}\varepsilon^{ab}_{~~cd}\left([\eta_{1}\stackrel{\star}{,}\Sigma^{cd}]+
[\omega^{cd}\stackrel{\star}{,}\lambda_{1}]\right)\right\}=0,
\label{ecomega}
\end{equation}
where we set $\omega^4=\eta_{1}+i\eta_{2}$ and $\Sigma^4=
\lambda_{1}+ i \lambda_{2}$.

Considering the expansion in powers of $\theta$ of the fields
(\ref{serie}) and of the Moyal product, expanding order by order
we get for the zero-th order
\begin{equation}
d\Sigma^{(0)ab}-\omega^{(0)ac} \wedge \Sigma^{(0)b}_c +
\omega^{(0)bc} \wedge \Sigma^{(0)a}_c=0,
\end{equation}
which is the second Cartan's structure equation for
$\omega^{(0)}$, with the solution given by (\ref{conexion}). To
first order in the $\theta$-expansion we have that (\ref{ecomega})
yields
\begin{eqnarray}
\theta^{\alpha\beta}\bigg\{&-&
[\omega^{(1)}_{\alpha\beta},\Sigma^{(0)}]^{ab}
+\frac{1}{2}\bigg[\partial_\alpha\eta_{2}^{(0)}\wedge\partial_\beta\Sigma^{(0)ab}
 + \partial_\alpha\omega^{(0)ab}\wedge\partial_\beta\lambda_{2}^{(0)}
\nonumber\\ &+&\frac{1}{2}\varepsilon^{ab}_{~~cd}\left(\partial_\alpha\eta_{1}^{(0)}\wedge\partial_\beta\Sigma^{(0)cd}
 +\partial_\alpha\omega^{(0)cd}\wedge\partial_\beta\lambda_{1}^{(0)}\right)\bigg]\bigg\}=0,
\end{eqnarray}
where $\Sigma^{(0)ab}=e^a\wedge e^b$ and
$\omega^{(0)ab}$ is given by (\ref{conexion}).

Solving for $\omega^{(1)}$, we have after some computations
\begin{eqnarray}
&&
e\left(\varepsilon^{abcd} \omega_{\alpha\beta,dc}^{(1)~~~e}-
\varepsilon^{abce}
\omega_{\alpha\beta,dc}^{(1)~~~d}\right)e_e^{~\sigma}
=\frac{1}{2}\varepsilon^{\mu\nu\rho\sigma} \bigg[
\partial_\alpha\eta_{2\mu}^{(0)}\partial_\beta\Sigma_{\nu\rho}^{(0)ab} +
\partial_\alpha\omega_\mu^{(0)ab}\partial_\beta\lambda_{2\nu\rho}^{(0)}
\nonumber \\&+& \left.\frac{1}{2}
\varepsilon^{ab}_{~~cd}\left(\partial_\alpha\eta_{1\mu}^{(0)}\partial_\beta
\Sigma_{\nu\rho}^{(0)cd} + \partial_\alpha \omega_\mu^{(0)cd}
\partial_\beta\lambda_{1\nu\rho}^{(0)}\right)-(\alpha\leftrightarrow\beta)\right]
\equiv M_{\alpha\beta}^{~~\sigma ab} (\Phi^{(0)}),
\end{eqnarray}
where $\Phi^{(0)}$ are real combinations of the tetrad $e^a$ and
the fields $\eta^{(k)}_{1}$, $\eta^{(k)}_{2}$, $\lambda^{(k)}_{1}$
and $\lambda^{(k)}_{2}$ for $k < n$. This equation can be
rewritten as
\begin{equation}
\omega_{\alpha\beta,ab}^{(1)~~~c}-\omega_{\alpha\beta,ba}^{(1)~~~c}+\omega_{\alpha\beta,da}^{(1)~~~d}
\delta_b^{ \ c} - \omega_{\alpha\beta,db}^{(1)~~~d}\delta_a^{ \ c}
= \varepsilon_{abde}e^{-1}M_{\alpha\beta}^{~~cde}(\Phi^{(0)}),
\end{equation}
from which we get
\begin{equation}
\omega^{(1)~~~c}_{\alpha\beta,ab} -
\omega^{(1)~~~c}_{\alpha\beta,ba}=\frac{1}{2}
(\varepsilon_{abde}\delta_f^{ \ c}+2\varepsilon_{abdf}\delta_e^{ \
 c})
M_{\alpha\beta}^{~~fde}={M}_{\alpha\beta,ab}^{(1)~~~c}(\Phi^{(0)})
\end{equation}
and then
\begin{equation}\label{conexion1}
\omega^{(1)}_{\alpha\beta,abc} =
\frac{1}{2}\bigg(M_{\alpha\beta,abc}^{(1)}-M_{\alpha\beta,bca}^{(1)}+M^{(1)}_{\alpha\beta,cab}\bigg).
\end{equation}
Thus $\omega^{(1)}_{\alpha\beta,abc}$ is determined by the tetrad,
$\eta^{(0)}_{1}$, $\eta^{(0)}_{2}$, $\lambda^{(0)}_{1}$ and
$\lambda^{(0)}_{2}$. Furthermore, to the $n$-th order we get from
(\ref{ecomega})
\begin{equation}
2d\Sigma_{\alpha_1\beta_1\cdots
\alpha_n\beta_n}^{(n)~~~~~~~~ab}-2[\omega_{\alpha_1\beta_1\ldots\alpha_n\beta_n}^{(n)},\Sigma^{(0)}]^{ab}
+M_{\alpha_1\beta_1\cdots
\alpha_n\beta_n}^{~~~~~~~~~~~ab}(\Phi^{(0)})=0,
\end{equation}
where $\Sigma^{(n)}$ vanishes if $n$ is odd and otherwise depends
on the tetrad by the ansatz (\ref{sigma}). Moreover, by a similar
computation as for the zero-th and first order cases we get
\begin{equation}
\omega^{(n)~~~~~~~~~~~~c}_{\alpha_1\beta_1\ldots\alpha_n\beta_n,ab}
-
\omega^{(n)~~~~~~~~~~~~c}_{\alpha_1\beta_1\ldots\alpha_n\beta_n,ba}
=
M_{\alpha_1\beta_1\ldots\alpha_n\beta_n,ab}^{(n)~~~~~~~~~~~c}(\Phi^{(n)}),
\end{equation}
from which the $n$-th correction to the spin
connection is given by
\begin{equation}
\omega_{\alpha_1\beta_1\ldots\alpha_n\beta_n,abc}^{(n)}
=\frac{1}{2}\bigg(M^{(n)}_{\alpha_1\beta_1\ldots\alpha_n\beta_n,abc}-
M^{(n)}_{\alpha_1\beta_1\ldots\alpha_n\beta_n,bca}+M^{(n)}_{\alpha_1\beta_1\ldots\alpha_n\beta_n,cab}\bigg).
\label{conexionn}
\end{equation}
Therefore, if we substitute the spin connection obtained to all
orders from these equations into the action (\ref{ncaccion}), we
get a noncommutative action for Einstein gravity, which depends to
all orders on $e_\mu^{~a}$, $\Sigma^4$ and $\omega^4$, in such a
way that to zero-th order it coincides with the Einstein-Hilbert
action. To first order action (\ref{ncaccion}) is given by
$$
A=i\int\bigg[\Sigma^{(0)ab(+)} \wedge
R^{(0)(+)}_{ab}-4\Sigma^{(0)4} \wedge d\omega^{(0)4}\bigg] +
i\theta^{\alpha\beta}\int\bigg[ -4\Sigma^{(1)4}_{\alpha\beta}
\wedge d\omega^{(0)4} -4\Sigma^{(0)4}\wedge
d\omega^{(1)4}_{\alpha\beta}
$$
$$
+\Sigma^{(0)ab(+)} \wedge \left(d\omega^{(1)(+)}_{\alpha\beta,
ab}-2\omega^{(0)c(+)}_{~~a} \wedge
\omega^{(1)(+)}_{\alpha\beta,cb}-\omega^{(0)4} \wedge
\omega^{(1)(+)}_{\alpha\beta,ab}+ \omega^{(0)(+)}_{ab} \wedge
\omega^{(1)4}_{\alpha\beta}\right)\bigg]
$$
$$
 -\frac{1}{2}\theta^{\alpha\beta}\int\bigg[\Sigma^{(0)ab(+)}
\wedge \bigg( -\partial_\alpha\omega^{(0)c(+)}_{~~a}\wedge
\partial_\beta\omega^{(0)(+)}_{cb} + 2\partial_\alpha\omega^{(0)(+)}_{ab}\wedge
\partial_\beta\omega^{(0)4}\bigg)
$$
\begin{equation}
-4\Sigma^{(0)4} \wedge \left(\partial_\alpha\omega^{(0)4} \wedge
\partial_\beta\omega^{(0)4}-
\frac{1}{4}\partial_\alpha\omega^{(0)ab(+)} \wedge
\partial_\beta\omega^{(0)(+)}_{ab} \right)\bigg]
+{\cal O}(\theta^2), \label{expansion}
\end{equation}
where $\Sigma^{(0)ab}=e^a\wedge e^b$ and $\omega^{(0)}_{ab}$ and
$\omega^{(1)}_{\alpha\beta,ab}$ are given by (\ref{conexion})
resp. (\ref{conexion1}).

It is worth to note that, simultaneously to the variation of
$\omega^{(+)ab}$ we can vary with respect to $\omega^4$ and
$\Sigma^4$, with the resulting equations of motion
\begin{equation}
\varepsilon^{\mu\nu\rho\sigma}\partial_\mu\Sigma_{\nu\rho}^4=
\varepsilon^{\mu\nu\rho\sigma}\left[\Sigma_{\mu\nu}^4\star\omega_\rho^4-
\omega^4_\rho\star \Sigma_{\mu\nu}^4+\frac{1}{4}\left(
\omega_{\mu, ab}^{(+)}\star\Sigma_{\nu\rho}^{ab(+)}-
\Sigma_{\nu\rho}^{ab(+)}\star\omega_{\mu, ab}^{(+)}\right)\right]
\end{equation}
and
\begin{equation}
\partial_\mu\omega_\nu^ 4-\partial_\nu\omega_\mu^ 4=
\frac{1}{4}\left(\omega_\mu^{ab(+)}\star\omega_{\nu ab}^{(+)}-
\omega_\nu^{ab(+)}\star\omega_{\mu ab}^{(+)}\right)
+\omega_\mu^4\star\omega^4_\nu -\omega_\nu^4\star\omega^4_\mu.
\end{equation}
To zero-th order we have the equations
$\varepsilon^{\mu\nu\rho\sigma}\partial_\nu\Sigma_{\nu\rho}^{(0)4}=0$
and
$\partial_\mu\omega_\nu^{(0)4}-\partial_\nu\omega_\mu^{(0)4}=0$,
which have the solutions
\begin{equation}
\Sigma_{\mu\nu}^{(0)4}=\partial_\mu S_\nu-\partial_\nu S_\mu, \ \
\ \ \ \ \ \omega_\mu^{(0)4}=\partial_\mu\phi.
\label{omega0}
\end{equation}
To higher orders the equations are of the form
\begin{eqnarray}
\varepsilon^{\mu\nu\rho\sigma}\partial_\nu\Sigma_{\rho\sigma,\alpha_1\beta_1\ldots\alpha_n\beta_n}^{(n)4}&=&
{\rm function~of}\ \Sigma^{(k)4 },
~\omega^{(k)4},~\omega_{ab}^{(k)},~ \Sigma_{ab}^{(k)} ~ \ \ \ {\rm
for} \ \ (k<n),
\nonumber\\
\partial_\mu\omega^{4(n)}_{\nu,\alpha_1\beta_1\ldots\alpha_n\beta_n}-
\partial_\nu\omega^{4(n)}_{\mu,\alpha_1\beta_1\dots\alpha_n\beta_n}&=&
{\rm function~ of}\ \omega_\mu^{(k)ab}, ~ \omega_\mu^{(k)4}~ \ \ \
{\rm for} \ \ (k<n).
\end{eqnarray}
Thus these equations, together with the equations of
$\omega^{ab}_\mu$, could be solved recursively.

\section{Final Remarks}

In the present paper we pursue the idea of the implementation of
the twisted symmetries to describe a non-commutative theory of
gravity. We applied the prescription based in the twisted gauge
transformations to construct a noncommutative gauge theory of
gravitation. In particular, we study noncommutative Pleba\'nski's
self-dual gravity. As well known it is a topological constrained
$SL(2,\mathbb{C})$ BF theory \cite{pleban}. This is addressed by
extending the fields to the universal enveloping albegra of
$su(2)$, given by $u(2)$. This action is constructed to be
invariant under twisted Lorentz and twisted diffeomorphism
transformations. The constraints are implemented at the
noncommutative level by the ansatz (\ref{sigma}). This ansatz
allows to solve the resulting torsion constraint to every order in
the expansion of the noncommutative parameter $\theta$ (see Eq.
(\ref{conexionn})). It is shown that at any order, the solution is
described in terms of the tetrad and the extra fields
corresponding to the fourth components of the connection $\omega$
and of the $B$-field two-form $\Sigma$, due to the enveloping
algebra. Furthermore, the noncommutative $BF$ action is explicitly
obtained to first order in $\theta$ (\ref{expansion}). It is
important to remark that, although the $BF$ theory is invariant
under twisted diffeomorphisms, the invariance of the resulting
noncommutative gravity theory is realized not directly through
metric variables as it was described at \cite{wessone}, but by
means of $\Sigma$ and $\omega$, through the prescription given in
Refs. \cite{Vassilevich:2006tc,Aschieri:2006ye} for gauge
theories. Then twisted diffeomorphisms are encoded in the twisted
gauge symmetry.

Finally, it is worth to mention that this procedure can be carried
over to define the classical topological invariants arising in
topological gravity \cite{topo}, in a way invariant also under
twisted diffeomorphisms. This issue was not enough clear in that
paper \cite{topo} and with these methods it can be clarified. Some
of results on this subject will be reported elsewhere.

\vskip 2truecm
\centerline{\bf Acknowledgments}

This work was supported in part by CONACyT M\'exico Grants 45713-F
and 51306 and also by projects PROMEP-UGTO-CA3 and VIEP-08/EXC/07.
The work of S.E.-J. was supported by a CONACyT postdoctoral
fellowship.

\vskip 2truecm
\appendix
\section{}

Let us consider a linear operator ${\cal O}$ acting on a set of
generic fields $\phi_i$. This operator acts locally as a matrix as
well as linearly in the derivatives of the field, and is defined
by
\begin{equation}
S^\phi_{\cal O}\phi=\int dz [{\cal O}^{\phi}(z) \phi(z)]_i
\frac{\delta}{\delta \phi_i(z)},
\end{equation}
where
\begin{equation}
[{\cal O}^{\phi}(x) \phi(x)]_i=O^{(1)}_{ij}(x) \phi_j(x)+
O^{(2)\mu}_{ij}(x)\partial_\mu \phi_j(x)={\cal O}^{A}(x) {\cal
T}^{\phi}_A \phi(x).
\label{o1}
\end{equation}
Here the operators ${\cal T}^{\phi}_A$ are constant and contain
the matrix and the differential actions on the fields $\phi_i$.

The coproduct of this operator is given by
\begin{equation}
\Delta(S_{\cal O})(\phi\otimes\psi)= (S_{\cal O}^{\phi}\otimes
1+1\otimes S_{\cal O}^{\psi})(\phi\otimes\psi).
\end{equation}

Let us now define the noncommutative coproduct as
\begin{equation}
\delta^\star(\phi \star \psi)=
\mu_\star[\Delta_{\theta}(\phi\otimes\psi)],
\end{equation}
where $\Delta_\theta\equiv \Delta_{\cal F}= {\cal
F}^{-1}\Delta(S^\phi_{\cal O}){\cal F}$ with ${\cal F}$ given by
Eq. (\ref{twist}).

In order to compute it, we must expand the exponentials. The
action of ${\cal F}$ gives the Moyal product on the fields $\phi$
and $\psi$, hence
\begin{eqnarray}
&&\delta^\star(\phi_i\star\psi_k)= \sum_n
\frac{1}{n!}\left(-\frac{i}{2}\right)^n
\theta^{\mu_1\nu_1}\cdots\theta^{\mu_n\nu_n}\nonumber\\
&& \times \mu_\star \left\{{\cal F}^{-1} (S_{\cal O}^{\phi}\otimes
1+1\otimes S_{\cal O}^{\psi})
\left[\partial_{\mu_1}\cdots\partial_{\mu_n}\phi_i(x)
\otimes\partial_{\nu_1}\cdots\partial_{\nu_n}\psi_k(x)\right]\right\}.
\end{eqnarray}
The action of $S_{\cal O}$ on the derivatives of the fields can be
computed as follows
$$
S^\phi_{\cal
O}\big[\partial_{\mu_1}\cdots\partial_{\mu_n}\phi_i(x)\big]= \int
dz [{\cal O}^{\phi}(z) \phi(z)]_j \frac{\delta}{\delta \phi_j(z)}
\partial_{\mu_1}\cdots\partial_{\mu_n}\phi_i(x)
$$
\begin{equation}
=\partial_{\mu_1}\cdots\partial_{\mu_n}[{\cal O}^{\phi}(x)
\phi(x)]_i.
\end{equation}
Consequently, we have
$$
\delta^\star(\phi_i\star\psi_k)=\sum_n
\frac{1}{n!}\left(-\frac{i}{2}\right)^n
\theta^{\mu_1\nu_1}\cdots\theta^{\mu_n\nu_n}
$$
\begin{equation}
\times \mu_\star\bigg\{{\cal F}^{-1}\bigg(
\partial_{\mu_1}\cdots\partial_{\mu_n}[{\cal O}^\phi(x)\phi(x)]_i
\otimes\partial_{\nu_1}\cdots\partial_{\nu_n}\psi_k(x)+\partial_{\mu_1}\cdots\partial_{\mu_n}\phi_i(x)
\otimes\partial_{\nu_1}\cdots\partial_{\nu_n}[{\cal
O}^\psi(x)\psi(x)]_k\bigg)\bigg\}.
\label{c1}
\end{equation}
The action of ${\cal F}^{-1}$ on the first term on the r.h.s. of
(\ref{c1}) can be written as follows
\begin{eqnarray}
&&\sum_m \frac{1}{m!}\left(\frac{i}{2}\right)^m
\theta^{\rho_1\sigma_1}\cdots
\theta^{\rho_m\sigma_m}\nonumber\\
& \times \bigg\{& \int dz_1 \partial_{\rho_1}\phi_{j_1}(z_1)
\frac{\delta}{\delta \phi_{j_1}(z_1)}\cdots \int dz_m
\partial_{\rho_m}\phi_{j_m}(z_m) \frac{\delta}{\delta
\phi_{j_m}(z_m)}
\partial^x_{\mu_1}\cdots\partial^x_{\mu_n}[{\cal O}^\phi(x)\phi(x)]_i\nonumber\\
&\otimes&\int dy_1 \partial_{\sigma_1}\psi_{l_1}(y_1)
\frac{\delta}{\delta \psi_{l_1}(y_1)}\cdots \int dy_m
\partial_{\sigma_m}\psi_{l_m}(y_m) \frac{\delta}{\delta
\psi_{l_m}(y_m)}
\partial^x_{\nu_1}\cdots\partial^x_{\nu_n}\psi_k(x)\bigg\}.
\end{eqnarray}
Furthermore, taking into account the definition (\ref{o1}), the
terms inside the biggest bracket in the preceding expression can
be written as
\begin{eqnarray}
&&\partial^x_{\mu_1}\cdots\partial^x_{\mu_n}{\cal O}^{\phi A}(x)
\int dz_1 \partial_{\rho_1}\phi_{j_1}(z_1) \frac{\delta}{\delta
\phi_{j_1}(z_1)}\cdots \int dz_m \partial_{\rho_m}\phi_{j_m}(z_m)
\frac{\delta}{\delta \phi_{j_m}(z_m)}
[{\cal T}_A\phi(x)]_i\nonumber\\
&\otimes&\partial^x_{\nu_1}\cdots\partial^x_{\nu_n} \int dy_1
\partial_{\sigma_1}\psi_{l_1}(y_1) \frac{\delta}{\delta
\psi_{l_1}(y_1)}\cdots \int dy_m
\partial_{\sigma_m}\psi_{l_m}(y_m) \frac{\delta}{\delta
\psi_{l_m}(y_m)} \psi_k(x).\label{t1}
\end{eqnarray}
From the properties of the Dirac function, considering that the
operators ${\cal T}_A$ in general are constant matrices and
contain derivatives, we have
\begin{equation}
\int dz_m \partial_{\rho_m}\phi_{j_m}(z_m) \frac{\delta}{\delta
\phi_{j_m}(z_m)} [{\cal T}_A\phi(x)]_i=[{\cal
T}_A\partial_{\rho_m}\phi(x)]_i=
\partial_{\rho_m}[{\cal T}_A\phi(x)]_i,
\label{vgaa}
\end{equation}
because ${\cal T}_A$ commutes with the derivatives. Therefore we
have from (\ref{vgaa}) the following
\begin{eqnarray}
&&\sum_m \frac{1}{m!}\left(\frac{i}{2}\right)^m
\theta^{\rho_1\sigma_1}\cdots
\theta^{\rho_m\sigma_m}\nonumber\\
&\bigg(&\partial_{\mu_1}\cdots\partial_{\mu_n}\big\{{\cal O}^{\phi
A}(x)
\partial_{\rho_1}\cdots \partial_{\rho_m}[{\cal T}_A\phi(x)]_i\big\}
\otimes\partial_{\nu_1}\cdots\partial_{\nu_n}
\partial_{\sigma_1}\cdots\partial_{\sigma_m}\psi_k(x)\nonumber\\
&+&\partial_{\mu_1}\cdots\partial_{\mu_n}\partial_{\rho_1}\cdots
\partial_{\rho_m}\phi(x)_i
\otimes\partial_{\nu_1}\cdots\partial_{\nu_n}\big\{{\cal O}^{\phi
A}(x)
\partial_{\sigma_1}\cdots\partial_{\sigma_m}[{\cal T}_A\psi(x)]_k\big\}
\bigg).\label{t2}
\end{eqnarray}
Inserting this expression back into (\ref{c1}) and considering
that the sum over $n$ gives ${\cal F}$, which compensates ${\cal
F}^{-1}$, we get
$$
\delta^\star(\phi_i\star\psi_k)= \sum_m
\frac{1}{m!}\left(\frac{i}{2}\right)^m
\theta^{\rho_1\sigma_1}\cdots \theta^{\rho_m\sigma_m}
$$
\begin{equation} \times \bigg\{{\cal O}^{\phi A}(x)
\partial_{\rho_1}\cdots \partial_{\rho_m}[{\cal T}_A\phi(x)]_i
\otimes\partial_{\sigma_1}\cdots\partial_{\sigma_m}\psi_k(x)
+\partial_{\rho_1}\cdots \partial_{\rho_m}\phi(x)_i \otimes{\cal
O}^{\psi A}(x)\partial_{\sigma_1}\cdots\partial_{\sigma_m} [{\cal
T}_A\psi(x)]_k\bigg\}.
\label{t3}
\end{equation}
Then we have
\begin{equation}
\Delta_\theta(S_{\cal O})[\phi_i(x)\otimes\psi_k(x)]= [{\cal
O}^{\phi}(x)\otimes1+1\otimes{\cal O}^{\psi}(x)]
[\phi_i(x)\otimes\psi_k(x)].\label{t4}
\end{equation}
From the above computation one can conclude that
$$
\delta^\star(\phi_i \star \psi_k) = \mu_\star\bigg[ {\cal
O}^\phi(x) \phi_i(x) \otimes \psi_k + \phi_i \otimes  {\cal
O}^\psi(x) \psi_k \bigg]
$$
\begin{equation}
= {\cal O}^\phi(x) \phi_i(x) \star \psi_k + \phi_i \star {\cal
O}^\psi(x) \psi_k.
\end{equation}
For instance, if we consider the gauge transformations (\ref{g1})
and translations on scalar fields (\ref{tr1}), then we get
correspondingly
\begin{equation}
\delta^\star_\alpha(\phi\star\psi)={\alpha}^l \cdot
(T_l\phi\star\psi+\phi\star T_l\psi) \label{t4}
\end{equation}
and
\begin{equation}
\delta^\star_\xi(\phi\star\psi)=-{\xi}^\mu \cdot
(\partial_\mu\phi\star\psi+ \phi\star\partial_\mu\psi)=-{\xi}^\mu
\cdot
\partial_\mu(\phi\star\psi). \label{t4}
\end{equation}

\section{}
Let us consider a noncommutative BF theory in four dimensions
(without cosmological constant term) with gauge algebra $su(2$).
The action is given by $I=\int {\rm Tr}B\wedge\widehat F$, where
the gauge field is $A=A^iT_i$ (with $T_i$ being the $su(2)$
generators) whose field strength is $\widehat{F}= dA + A
\stackrel{\star}{\wedge} A = (dA^i + i \varepsilon_{~jk}^{i}A^j
\stackrel{\star}{\wedge} A^k)T_i + A^i\stackrel{\star}{\wedge}A_i$
and $B=B^iT_i$ is a two-form field. This action is invariant under
twisted $su(2)$ gauge transformations \cite{Vassilevich:2006tc} as
it has been shown in our Section 2. Moreover, due to the trace
keeps only the $su(2)$ part of the field strength we get
\begin{equation}
I=\int B^i\wedge(dA_i+i\varepsilon_{ijk}A^j\stackrel{\star}{\wedge}A^k).
\label{b1}
\end{equation}
However, as shown in \cite{Aschieri:2006ye} for Yang-Mills theory,
the consistency of the equations of motion requires the enveloping
algebra. In this appendix we argue that it is also the same
situation for BF actions in the case of $su(2)$.

The field equations of the action (\ref{b1}) are given by
\begin{eqnarray}
\varepsilon^{\mu\nu\rho\sigma}\left(\partial_\mu B_{\nu\rho}^i
-\frac{i}{2}\varepsilon^i_{~jk}\{A_\mu^j\stackrel{\star}{,}B_{\nu\rho}^k\}\right)&=&0,
\label{b2}\\
\partial_\mu A^i_\nu-\partial_\nu A^i_\mu+ i\varepsilon^i_{~jk}
\{A_\mu^j\stackrel{\star}{,}A_{\nu}^k\}&=&0.\label{b3}
\end{eqnarray}
The integrability conditions of the first equations are
\begin{equation}
\varepsilon^{\mu\nu\rho\sigma}\varepsilon^i_{~jk}\partial_\mu
\{A_\nu^j\stackrel{\star}{,}B_{\rho\sigma}^k\}=
\varepsilon^{\mu\nu\rho\sigma}\varepsilon^i_{~jk}\left(
\{\partial_\mu A_\nu^j\stackrel{\star}{,}B_{\rho\sigma}^k\}+
\{A_\nu^j\stackrel{\star}{,}\partial_\mu B_{\rho\sigma}^k\}
\right)=0.
\end{equation}
However, if we use the equations (\ref{b2}) and (\ref{b3}), we
have, after some manipulations
\begin{equation}
\varepsilon^{\mu\nu\rho\sigma}\varepsilon^i_{~jk}\partial_\mu
\{A_\nu^j\stackrel{\star}{,}B_{\rho\sigma}^k\}=
-\frac{i}{2}\varepsilon^{\mu\nu\rho\sigma}\left[
\{A_\mu^i\stackrel{\star}{,}\{A_\nu^j\stackrel{\star}{,}B_{\rho\sigma
j}\}\} +\{A_\mu^j\stackrel{\star}{,}\left(\{A_\nu^i,B_{\rho\sigma
j}\}- \{A_{\nu j},B_{\rho\sigma}^i\}\right)\right],
\end{equation}
which does not vanish identically. If instead of $su(2)$, we had
considered the enveloping algebra $u(2)$, the corresponding
equations would vanish due to the generalized Jacobi identities.
Hence, even if the action is invariant under any Lie algebra, the
consistency of the equations of motion requires the whole
enveloping algebra.

\vskip 2truecm

\end{document}